\documentclass{aa}
\usepackage[below]{placeins}
\usepackage{amsfonts}
\usepackage{graphicx,epsfig}
\usepackage{natbib}

\newcommand\R{\mathbb{R}}

\begin{document}
\title{Robust Reconstruction from Chopped and Nodded Images}

\author{
Frank Lenzen\inst{1}, Otmar Scherzer\inst{1}
\and
Sabine Schindler \inst{2}}
\authorrunning{F. Lenzen et al.}
\institute{ 
Institute of Computer Science, University of Innsbruck,
Technikerstra\ss e 21a, 6020 Innsbruck, Austria
\and
Institute for Astrophysics, University of Innsbruck,
Technikerstra\ss e 25, 6020 Innsbruck, Austria}
\offprints{Frank Lenzen, \email{Frank.Lenzen@uibk.ac.at}}
\date{Received May 11, 2005  / Accepted July 30, 2005}
\abstract{
In ground based infrared imaging a well-known technique to reduce the influence of 
thermal and background noise is \emph{chopping and nodding}, where four different signals of 
the same object are recorded from which the object is reconstructed numerically. 

Since noise in the data can severely affect the reconstruction, regularization
algorithms have to be implemented.

In this paper we propose to combine iterative reconstruction algorithms with robust 
statistical methods. 

Moreover, we study the use of multiple chopped data sets with different chopping 
amplitudes and the according numerical reconstruction algorithm. Numerical simulations 
show robustness of the proposed methods with respect to noisy data.

}
\maketitle

%----------------------------

\section{Introduction}
%----------------------------
Data collected in infrared ground based astronomy are affected by atmospheric and telescopic 
thermal background noise. A common approach for noise reduction is chopping and nodding 
\citep{EMERSON,ROBBERTO}. \\

In \emph{chopping} the secondary mirror of the telescope is moved 
(cf. \citet{BERTERO03a}) and signals are recorded for two different tilt angles. We denote 
the positions before and after tilting by $A$ and $B=A+h$, and the recorded signals by 
$S_1$ and $S_2$, respectively. Afterwards, in a \emph{nodding} procedure the telescope is 
moved to the position $C=A-h$ and the signal $S_3$ is recorded. 
Then a second chopping procedure is applied and a signal $\tilde{S}_1$ is recorded 
at position $A$.

In literature the 
vector $h$ is referred to as the chopping amplitude.
The data visualized after chopping and nodding is 
\[
f=S_1-S_2-S_3+ \tilde{S}_1
\]
and the image $u$ to be reconstructed satisfies
 \[
2 u(.) -u(.-h)-u(.+h)=f(.)\;.
\]
When observing a point-like and isolated object the chopping amplitude can be chosen in such way 
that at positions $B$ and $C$ only empty sky is recorded in which case the numerical 
reconstruction algorithm is dispensable. However, in practice the objects observed are often extended or other objects are nearby. In these cases the reconstruction step is mandatory.

Reconstruction from chopped and nodded data was first discussed in \citet{BECKERS}, where a 
Fourier-based reconstruction has been suggested.

In \citet{BERTERO99,BERTERO00,BERTERO03a, BERTERO03b} iterative reconstruction algorithms 
have been investigated. Recently, in \citet{CHAN} reconstruction algorithms based on wavelet 
decomposition have been proposed.
 
Experiments have shown that noise can severely 
affect the reconstruction process \citep[see][]{KAEUFL}. 
In this paper we investigate two different approaches for robust reconstruction in the 
presence of a significant amount of noise.

The outline of this paper is as follows: in Sect.~\ref{SectProblem} we derive the model 
of chopping and nodding and review the reconstruction algorithms 
introduced by \citet[cf.][]{BERTERO03a, BERTERO03b}.

In this paper we use iterative regularization techniques as 
well and thus Bertero's work is the most relevant to compare with. However for the 
sake of noise robustness we combine iterative regularization techniques with \emph{robust} 
filtering methods from statistics (cf. Sect.~\ref{SectRobust}).
Moreover, we study the combined reconstruction from multiple chopped data 
with different chopping amplitudes and the 
effect on the quality of the reconstruction. 
Finally, in Sect.~\ref{SectResults} we present some numerical experiments. 
%The paper ends with a conclusion, cf. Sect.~\ref{SectConclusion}.

%----------------------------
\section{Problem description and basic reconstruction methods}\label{SectProblem}
%----------------------------
\subsection{Continuous problem}

The problem description basically follows \cite{BERTERO03a, BERTERO03b}, but differs 
in the way that we take into account two-dimensional chopping amplitudes, while in 
\cite{BERTERO03a, BERTERO03b} the data are preprocessed by appropriate rotations so 
that the chopping amplitude is along the principal axis of the data.

We denote by $u:\R^2\to\R$ the brightness intensity distribution in the sky, which 
we assume to be non-negative.

Let $\Omega=[0,l_x]\times[0,l_y]\subset \R^2$ be a section of the sky under 
investigation and let 
$h=\left(h_x,h_y\right)\not=0,\; h_x\ge 0, h_y\ge 0$, be the chopping amplitude. 
We define the operator
\begin{eqnarray}
&&I_h: L^2(\Omega_h)\to L^2(\Omega)\nonumber\\
&&u \to I_h(u)(x,y):=2 u(x,y)-u(x+h_x,y+h_y)\label{EqDefI}\\
&&\quad\quad\quad\quad-u(x-h_x,y-h_y),
\end{eqnarray} 
where $Y:=L^2(\Omega)$ and $X:=L^2(\Omega_h)$ are the spaces of square integrable 
functions on $\Omega$ and 
\begin{eqnarray*}
\Omega_h:=\Omega \cup\{(x,y)| &&(x+h_x,y+h_y)\in\Omega\mbox{ or }\\
&&(x-h_x,y-h_y)\in\Omega\},
\end{eqnarray*}
respectively.

The problem of reconstruction from chopped data $f(x,y):\Omega\to\R$ can be written 
as an operator equation
\begin {equation}
I_h(u)=f\label{EqOp}.
\end{equation}

A solution of (\ref{EqOp}) is also a minimizer of the functional 
\[
u \to \|I_h(u)-f\|^2\;,
\]
and therefore satisfies the corresponding optimality condition
\begin{equation}
I_h^*(I_h u-f)=0\label{NormEq}\;.
\end{equation}
$I^*_h$ is the adjoint operator to $I_h$, which satisfies 
\[
\int_{\Omega_h} I_h (u) v = \int_\Omega u I_h^*(v)  \mbox{ for all } u \in L^2(\Omega_h), 
v \in L^2(\Omega)\;.
\]
Since $I_h^*$ is injective, for any injective operator $J:X \to Y$, 
(\ref{EqOp}), (\ref{NormEq}) and 
\begin{equation}
J(I_h u-f)=0 \label{EqLavr}
\end{equation}
have the same solutions.  
Basic numerical methods for solving (\ref{EqLavr}) are based on fixed point 
iterations: 
\begin{equation}
\label{eq:Bert-A}
u^{i+1}:=u^{i}- \tau\; J(I_h u^i -f)\,,\quad u^0:=0,
\end{equation}
where $\tau$ is a relaxation parameter. For $J=I_h^*$ the fixed point iteration 
is commonly referred to as \emph{Landweber iteration}
\begin{equation}
\label{eq:Bert-B}
u^{i+1}:=u^{i}- \tau\; I_h^*(I_h u^i -f)\;,\quad u^0:=0.
\end{equation}

In \cite{BERTERO03a}, (\ref{eq:Bert-A}) has been implemented with an operator $J$ 
that extends $u$ periodically to $\Omega_h$. 

Here, as in \cite{BERTERO03b}, we use the operator 
\[
J(u)(x):=\left\{\begin{array}{ll}
u(x)& \mbox{ if } x\in \Omega\\
0 & \mbox{ else }
\end{array}
\right.,
\] 
which extends a function $u \in L^2(\Omega)$ to zero in $\Omega_h\setminus \Omega$.

\subsection{Discretization}
The domain $\Omega$ is discretized by a quadratic grid with nodes 
$(x_i,y_j)\in\Omega, i=1\dots N,j=1\dots M$ and cell length $1$. 
The nodes coincide with the sampling points of $u$ and $f$, i.e.,  
$\vec{u}:=(u_{i,j})=(u(x_i,y_i))$ and 
$\vec{f}:=(f_{i,j})$ are the sampling data. 

With $\vec{u}$ a bilinear interpolating function is associated.
Assuming $u \equiv 0$ on $\Omega_h\setminus \Omega$ the resulting discretized 
system of (\ref{EqOp}) is 
\begin{equation}
A_h \vec{u} = \vec{f}\label{EqDiscrete}\,,
\end{equation}
where $A_h$ is an $NM \times NM$ matrix.
Details on the structure of matrix $A_h$ and properties of its eigenvalues 
can be found in \citet{BERTERO03b}.
Note that matrix $A_h$ is symmetric and positive definite and thus (\ref{EqDiscrete})
has a unique solution. 

Since the null-space of $I_h$ is not trivial, the choice of the 
extension of $u$ into $\Omega_h \backslash \Omega$ enforces a particular solution 
of $I_h$ to be calculated.

\subsection{Review on reconstruction methods\label{SectMethods}}
Three different methods have been proposed in the literature for reconstruction from 
chopped and nodded data: Fourier-based reconstruction method 
\citep[cf.][]{BECKERS,BERTERO03a}, iterative 
reconstructions \citep[cf.][]{BERTERO03a}, and a wavelet based approach \citep[cf.][]{CHAN}.
These methods are reviewed below:
\begin{enumerate}
\item 
For applying the Fourier-based reconstruction (cf. \citet{BERTERO03a}) it is assumed 
      that $h$ is in vertical direction and the solution of (\ref{EqOp}) is periodic across 
      $\partial \Omega$ in chopping direction. 
      In this case the image can be reconstructed for each column separately,
      the discretized linear system of $I_h$ becomes 
      $\tilde{A}_h u = f$ with a circulant matrix $\tilde{A}_h$ and Fourier 
      techniques can be used for the efficient numerical solution:
      applying the discrete Fourier transform to the linear system it becomes 
      \[ {\cal{F}}(\tilde{A}_h) {\cal{F}}u = {\cal{F}}f\]
      where 
      ${\cal{F}}(\tilde{A}_h) = D:=diag(d_1,\dots d_M)$ is diagonal, and thus 
      can be solved efficiently provided the diagonal matrix has full rank.
\item 
The constrained Landweber iteration \citep{BERTERO03a}
(referred to as method (A)) reads as follows:
\[
\begin{array}{l}
u^{(0)}=0\\
\mbox{for } n=1\dots N:\\
\quad\quad u^{(n)}=P_+\left[u^{(n-1)}+\tau\,A_h^T\left(f- A_h u^{(n-1)}\right)\right]
\end{array}
\]
where $P_+$ is the projection operator onto the set of non-negative vectors 
(each negative entry is set to zero), and $\tau$ is a positive relaxation parameter.\\

The projected Lavrentiev iteration (method (B)) is defined by
\[
\begin {array}{l}
u^{(0)}=0\\
\mbox{for } n=1\dots N:\\
\quad\quad u^{(n)}=P_+\left[u^{(n-1)}+\tau\left(f-A_h u^{(n-1)}\right)\right]\;.
\end{array}
\]
      Both iterative methods work reasonably efficient if the chopped and nodded 
      data are only distorted by a small amount of noise. In this case, the results are qualitatively comparable to those obtained with the Fourier methods. 
      However, these methods suffer from robustness with respect to high noise distortions.
\item Finally we review a wavelet approach proposed by \citet{CHAN}.
      Here chopping is interpreted as high-pass filtering. Supplementing 
      this filter by a low-pass and a subsequent high-pass filtering a tight frame wavelet 
      system for a multi-resolution analysis is obtained. The Landweber method is then 
      combined in a multi resolution framework and wavelet thresholding is applied
      \citep[cf.][]{DONOHO94}. Wavelet thresholding is used for denoising in 
      each iteration step. Moreover, a post-processing step for removing artifacts 
      in feature-less areas is proposed.
\end{enumerate}

%\FloatBarrier
%-----------------------------
\section{Robust reconstructions\label{SectRobust}}
%-----------------------------

\subsection{Modifications of the iterative methods}

Simulations with noisy data show that the methods described above have a tendency to 
introduce \emph{artificial structures} from the noise (cf. Sect. \ref{ResultsClassical}).

We propose to combine iterative reconstruction methods with a median 
filtering technique in each iteration step.
The median filtering \citep[cf. e.g.][]{PESTMAN}) is the method of choice, since it 
removes artifical structures \citep[cf.][]{SOILLE} appearing in each iteration step.
Moreover, the median can be implemented very efficiently.

The median filter is defined as follows:

For an odd number of values 
$\{v_1\dots v_{2n+1}\}$ in ascending order, the median is $v_n$. In median filtering 
each value $u_{i_0,j_0}$ is replaced by the median value of surrounding values 
$u_{i,j}$ with indices $i,j$ in a neighborhood of $i_0,j_0$. We take as neighborhood 
$(i,j)$ satisfying $|i-i_0|\le s$ and $|j-j_0|\le s$ with $s=1,2,3$. 
The median filter can be implemented very efficiently (cf. \citet{SOILLE} and \citet{SONKA}).

We investigate the following variant of method (A) defined by\\
{\tt
$u^{(0)}=0$\\
for  $n=1\dots N:$\\
\rule{1cm}{0cm}1. $ u^{(n)}=P_+\left(u^{(n-1)}+
\tau\;A_h^T\left(f-A_h u^{(n-1)}\right)\right)$\\
\rule{1cm}{0cm}2. Apply the median filter to $u^{(n)}$.\\
}
and the variant of method (B) is defined accordingly.\\

This strategy of combination of an iterative method with additional filtering after each 
iteration is analogous to the strategy of the wavelet based approach in \citet{CHAN}, 
where wavelet thresholding is used for additional filtering after each iteration of 
the Landweber method. 
%Thus in the wavelet approach there is a strong
% coupling of reconstruction method and denoising technique.
% A comparison of the different filter techniques is given in....

\subsection{Reconstruction based on the conjugate gradient method}

%A feasible strategy is to solve Eq.~(\ref{EqDiscrete}) directly. 
Since $A_h$ is symmetric and positive-definite, it can be solved with the conjugate 
gradient (cg) method \citep[cf.][]{HANKE}, which has faster convergence properties than the Landweber and Lavrentiev methods.
However, the fast convergences also makes the method more sensitive to noise, which we 
overcome again by applying after one iteration of the cg-method an additional median filtering 
step. The modified cg-method reads as follows:

{\tt
\noindent
$u^{(0)}=0$\\
for $n=1\dots N:$\\
\rule{1cm}{0cm} 1. Apply one step of the cg -method with\\
\rule{1cm}{0cm} initial vector $u^{(n-1)}$ \\ 
\rule{1cm}{0cm} 2. Denoting the result by $\tilde {u}^{(n)}$,\\
\rule{1cm}{0cm} we replace $\tilde {u}^{(n)}$ by $P_+\tilde {u}^{(n)}$ to meet the\\
\rule{1cm}{0cm} constraint of non-negativity.\\
\rule{1cm}{0cm} 3. We apply the median filter to $\tilde{u}^{(n)}$\\
\rule{1cm}{0cm} to achieve iterate $u^{(n)}$.\\
}

The constraint of non-negativity is necessary to avoid
artifacts in the reconstruction.

%\FloatBarrier
%--------------------------------------------------------
\subsection{Multiple chopped data sets\label{multi}}
%--------------------------------------------------------

For an improvement of the quality of the reconstruction \citet{BECKERS,BERTERO00} proposed to use multiple chopped and nodded data sets.
The reconstruction is performed on each data set independently and the results are combined in a post-processing step 
calculating  the pointwise mean resp. the median. 

In \citet{BERTERO00} this strategy is used mainly to avoid two kinds of artifacts,
first, ghosts from bright sources, 
%i.e. multiple images of an object arranged in multiple 
%distances of the chopping amplitude, 
and second regions, where faint structures are superimposed by negative counterparts of 
bright sources and consequently the reconstructed image becomes zero artifically.

In the following we take advantage of multiple data sets for a robust reconstruction in the presence of noise.

Let $h^k,k=1\dots K$ denote a set of chopping amplitudes and denote by $f^k$ the corresponding sampled data sets.

A reconstruction $u$ is a solution of the system
\begin{equation}
A_{h^k}u =f^k \; k=1,\ldots,K\;.\label{EqMulti}
\end{equation}

We solve this system with a blocked Landweber-Kacmarcz method 
(for more details on such methods cf. \cite{KALTNEUSCHERZ}):\\

{\tt
\noindent
Set $u^{(0)} = 0$\\ 
for $n=1\dots N:$ (iterationindex)\\
\rule{1cm}{0cm} 1. For each $k=1,\ldots,K$ perform\\
\rule{1cm}{0cm} two cg iterations starting from\\
\rule{1cm}{0cm} $(u^{(n-1)})\;.$\\
\rule{1cm}{0cm} The solution is denoted by $u^{(n)}_{k}$.\\
\rule{1cm}{0cm} 2. Calculate the median $u^ {(n)}$ from \\
\rule{1cm}{0cm} of $u_1^ {(n)},\dots,u_K^ {(n)}$ for noise removal.\\
}

Note that the median is calculated separately in each sampling point for the stack of images:
That is in step 2 of the above algorithm we set
\[
(u^{(n)})_{i,j}:=\mbox{median }\{ (u_{1})_{i,j},\dots,(u_{K})_{i,j}\}
\]
for each $i,j$.\\

We also tested an alternative combination, namely to first
detect outliers in $\{(u_{1})_{i,j},\dots,(u_{K})_{i,j}\}$ via the standard deviation and then 
calculating the mean ignoring these outliers, which gives fairly similar results as when 
using the median.

%-----------------------------
\section{Numerical Experiments\label{SectResults}}
%-----------------------------

\subsection{Simulation of the Noise Process\label{SectNoise}}
Realistic noisy test data, which are used for our numerical experiments, require
detailed knowledge on the noise process.
There are two different sources of noise, which we have to be taking into account, 
background emission and thermal (dark) noise which is aimed to be reduced with chopping and nodding, 
and noise occurring during the data recording process, which is Poisson noise (due to photon counting errors of a CCD detector) and Gaussian read out noise.

We assume that the noise affects each of the four recorded signals independently.
Let $\eta(u(x,y))$ denote the random variable, then the recorded signal reads as follows
\begin{equation}
\begin{array}{l}
\tilde{f}(x,y): =2\left(u(x,y)\right)+\eta(u(x,y))+\eta(u(x,y))\\
              \quad - u(x+h_x,y+h_y)-\eta( u(x+h_x,y+h_y))\\
              \quad  -u(x-h_x,y-h_y)-\eta(u(x-h_x,y-h_y))\;.
\end{array}\label{EqChoppingNoise}
\end{equation}
We write  $\tilde{f}(x,y):=f(x,y)+\tilde{\eta}(x,y)$ resp. 
$\tilde{f}_{i,j}:=f_{i,j}+\tilde{\eta}_{i,j}$, where $\tilde{\eta}(x,y)$ 
\begin{eqnarray*}
&&\tilde{\eta}(x,y)=\eta(u(x,y))+\eta(u(x,y))\\
&&-\eta(u(x+h_h,y+h_y))-\eta(u(x-h_x,y-h_y)).\\
\end{eqnarray*}
According to our considerations, the random variable $\eta(u)$ is the sum of a Gaussian 
white random $\eta_1=N(0,\sigma_1),\;\sigma_1\ge 0$ (the background emission and read out noise) 
and a Poisson noise $\eta_2(u)$: $\eta(u):= \eta_1(u)+\eta_2(u)$.

We use test images with intensities scaled to the range of $[0,1]$. These intensities 
are related to the number of photons counted by the CCD detector. An intensity $u$ corresponds 
to a number of $N \cdot u$ photon counts with some unknown factor $N>0$.
The Poisson distribution of the photon counts can be approximated by a Gaussian distribution 
with standard deviation $ \sqrt{N\cdot u}$.

Scaling this distribution to the range $[0,1]$ the intensities are Gaussian distributed
with standard deviation $\sqrt{u/N}$. Therefore, we can approximate $\eta_2$ with a 
Gaussian distribution with standard deviation $\sqrt{\sigma_2 u}$ and $\sigma_2=\frac{1}{N}$.

%\FloatBarrier

\subsection{Behavior of the 'classical' methods\label{ResultsClassical}}

\begin{figure}[h]\begin{center}
\includegraphics[width=0.45\textwidth]{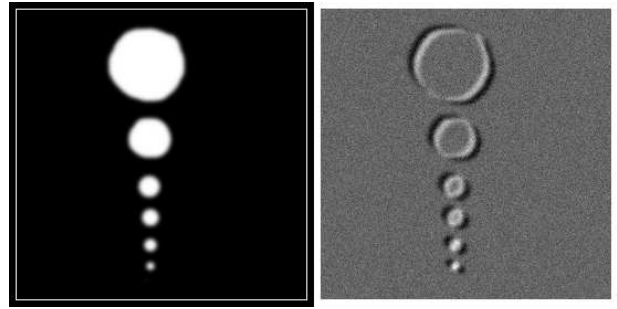}
\caption{\label{FigTestimage1}
Left: Test image. The white frame marks the domain $\Omega$, where the 
chopped data are recorded; right: Corresponding chopped data with chopping amplitude 
$h=(5,3)$ including noise with
variances $\sigma_1=0.05$ and $\sigma_2=0.0001$.}
\end{center}
\end{figure}

In the following we demonstrate that artificial structures appear in the reconstruction 
with 'classical' methods from the noise.\\

Hereby we concentrate on the case of chopping amplitudes being small with respect to the image size.\\

In the case of a horizontal chopping amplitude being an integer multiple 
of the pixel size the condition number of the matrix $A_h$ is of the order of 
the ratio of number of sampling points and the chopping amplitude 
\citep[cf.][]{BERTERO03b}.
Thus the problem becomes ill-conditioned for small chopping amplitudes 
and low-frequency eigenvectors are affected by noise in the data resulting 
producing the artificial structures \citep[cf.][]{BERTERO03b}.

%\begin{figure}[h]\begin{center}
%\includegraphics[width=0.3\textwidth]{IMAGES/figure4.eps}
%\caption{\label{FigTestimage1Noise}
%Chopped data from test image 1 with $h=(5,3)$ including noise with
%variances $\sigma_1=0.05$ and $\sigma_2=0.0001$.}
%\end{center}
%\end{figure}

Fig.~\ref{FigTestimage1} shows an artificial test image (left) and corresponding 
chopped and nodded data (right) with chopping amplitude $h=(5,3)$ which is distorted 

by two Gaussian noise processes with parameters $\sigma_1=0.05$ and $\sigma_2=0.0001$. 
The signal-to-noise-ratio is 25.4752
\footnote{Let $u:\Omega\to \R$ be a signal with minimum $u_{min}$ and maximum $u_{max}$, 
$\tilde{u}:\Omega\to\R$ a distorted signal and $\sigma$ the standard deviation of 
$u-\tilde{u}$, then the signal-to-noise-ratio is defined by 
$\mbox{SNR }:=\;20 \log_{10}\left(\frac{u_{max}-u_{min}}{\sigma}\right)$.}.

\begin{figure}[h]\begin{center}
\includegraphics[width=0.45\textwidth]{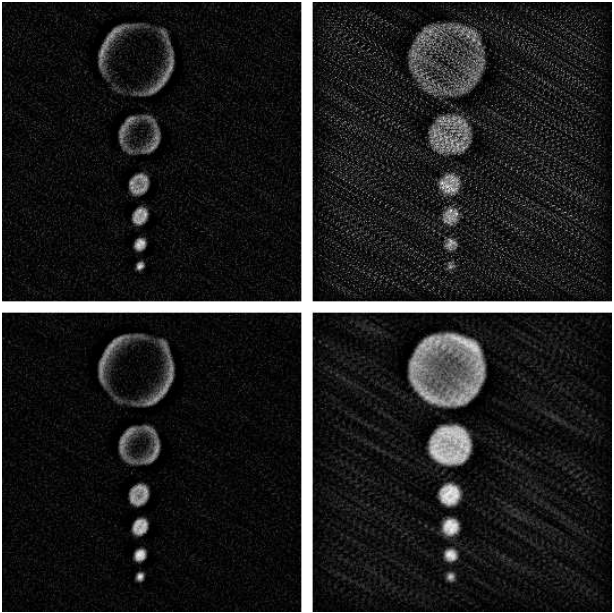}
\caption{\label{FigBertero1}
Top left: Reconstruction from the noisy data given in Fig.~\ref{FigTestimage1}, right, 
with method (B) after 10 iterations and $\tau=0.1$. 
Top right: Reconstruction with method (B) after 100 iterations and $\tau=0.1$.
Bottom left: Reconstruction with $h=(5.5,3.3)$ using method (B)
after 10 iterations and $\tau=0.1$. Bottom right: Reconstruction from the same data using 
Method (B) after 100 iterations and $\tau=0.1$.}
\end{center}
\end{figure}

Fig. \ref{FigBertero1}, top left and top right, resp., shows the reconstruction from these 
test data applying method (B) with $\tau=0.1$ for $h=(5,3)$ after 10 and 100 iterations, respectively.
We applied the same method to test data with a chopping amplitude $h=(5.5,3.3)$ not 
matching the grid spacing.
Reconstructions after 10 resp. 100 steps are shown in Fig.~\ref{FigBertero1}, bottom 
left resp. bottom right.

The  computation times \footnote{Computed on an AMD 64 FX 3500+, 
computations times have been averaged over several runs} of method B
were 0.05 seconds for $h=(5,3)$ and 10 steps, 0.2 seconds for $h=(5.5,3.3)$ and 10 steps, 
0.5  seconds for $h=(5,3)$ and 100 steps and 1.9 seconds for $h=(5.5,3.3)$ and 100 steps.

Since the chopped and nodded data $f$ provide information at the objects' edges,  
their reconstruction is satisfactory at an early stage of the iteration, whereas 
the objects' interior is reconstructed at a later stage. 
Therefore, larger objects require a larger number of 
iterations to be reconstructed satisfactory.

When the discretization points are aligned with the chopped discretization points the 
reconstruction is more efficient. If the points are not aligned, then 
the numerical reconstruction is smoother; moreover, the iterative algorithms are slower 
convergent, i.e. a larger number of iterations is needed for reconstructing the interiors
of the objects, which in turn leads to an amplification of noise and appearance of artificial structures.
Method (A), the cg-method, and the Fourier method shows similar convergence properties.
Applying the cg-based method to noisy data shows that the residual 
$\|A_hu-f\|$ decreases in the beginning but after some iterations is starts oscillating 
or is even increasing.
This suggests to stop the iteration of the cg-solver when the residual reaches its first local minimum.\\

Since Eq. (\ref{EqOp}) is linear we have $I_h(u_f+u_\eta)=f+\eta$, where 
$u_\eta$ are the data reconstructed from noise and $u_f$ is the reconstruction from 
noise free data.

For illustration of the noise amplification behavior of this equation, 
we assume that the data $f\equiv 0$ are disturbed at one grid point: 

\[
\eta_{i,j}=\left\{
\begin{array}{ll}
1 &\mbox{ if } i=i_0 \mbox{ and } j=j_0\\
0 &\mbox{else}
\end{array}
\right.
\]
for a  fixed $i_0\in\{1\dots N\}$, $j_0\in\{1\dots M\}$.

First, let us consider the Fourier-based reconstruction from this particular $\eta$ for horizontal chopping 
amplitudes.
In order to handle small eigenvalues of matrix $D$, we use a slightly modified method
replacing matrix $D$ by 
\[
D_\varepsilon^\dagger:=diag(d^\dagger_1\dots d^\dagger_N)
\]
where $\varepsilon>0$ is a small parameter and 
\[
d^ \dagger_i=
\left\{
\begin{array}{cc}
1/d_i& \mbox{ if } |d_i|>\varepsilon\\
\varepsilon&\mbox{ else}
\end{array}\right.,\quad i=1\dots N,
\]
%Note that this modification is not required assuming that the chopping amplitude
%$h$ is not a divisor of the image width. Nevertheless the modified method shows the same behaviour
%as the iterative methods, i.e. cutting off small eigenvalues of $A_h$.

\begin{figure}[h]\begin{center}
\includegraphics[width=0.4\textwidth]{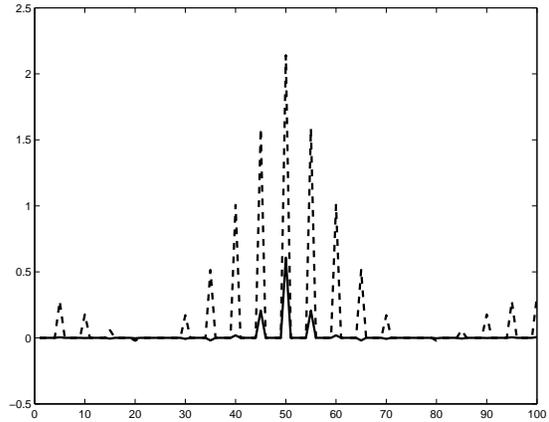}
\caption{\label{FigFourierReconstr} Horizontal cut of the image reconstructed with the Fourier method 
with $D_{0.1}$ (solid line) and $D_1$ (dashed line). 
The data are $f_{i,j}:=\delta_{i_0i}\delta_{j_0j}$ for afixed $i_0\in\{1\dots N\}$,$j_0\in\{1\dots M\}$ and the 
chopping amplitude is $h=(5,0)$.}
\end{center}
\end{figure}

Fig. \ref{FigFourierReconstr} shows a horizontal cut through point $(i_0,j_0)$ of the function
 reconstructed from $\eta$ for $h=(5,0)$  and $\varepsilon=0.1$ resp. $\varepsilon=1$.

\begin{figure}[h]\begin{center}
\includegraphics[width=0.45\textwidth]{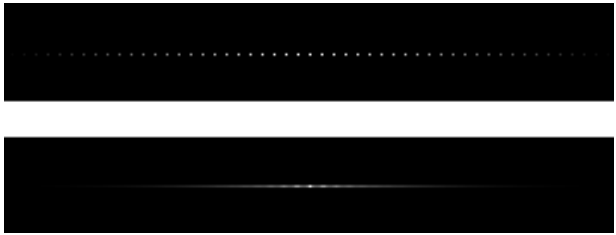}
\caption{\label{FigDelta}
Reconstruction from data $f_{ij}=\{ 1 \mbox{ if } i=i_0\mbox{ and } j=j_0, 0\mbox{ else }\}$, $i_0$ and $j_0$ fixed, 
using the cg-based method for $h=(5,0)$ (top) and $h=(5.5,0)$ (bottom).}
\end{center}
\end{figure}

Similar artificial structures can be observed in the reconstruction from iterative methods.
Fig. \ref{FigDelta} shows the reconstruction from data  $f$ for $h=(5,0)$ (top) and $h=(5.5,0)$ (bottom).

In the later case, as  
the unknowns of $u$ are far more coupled the reconstruction is smooth and thus more 
complicated to handle numerically.

\FloatBarrier

%---------------------------------------------------------------
\subsection{Results of the modified methods\label{SectResultsMod}}
%--------------------------------------------------------------
In this section we present numerical results of the modified methods applied on noisy data.

Firstly we used the Fourier method with inverse $D_{\varepsilon}$ to reconstruct our first 
test image. \footnote{ For implementation the FFTW-C-library \citep[][ resp. www.fttw.org]{FFTW} was used.}

But numerical experiments show that a small $\varepsilon$ is needed for a good reconstruction
for noise free data. 
Since the eigenvalues of the inverse tend to infinity for $\varepsilon\to 0$, 
artificial structures due to noise are significant for small 
$\varepsilon$.
For noisy data, a compromise between good reconstruction and little noise amplification is 
not possible. The results are not satisfactory, so they are not presented here.\\

\begin{figure}[h]\begin{center}
\includegraphics[width=0.45\textwidth]{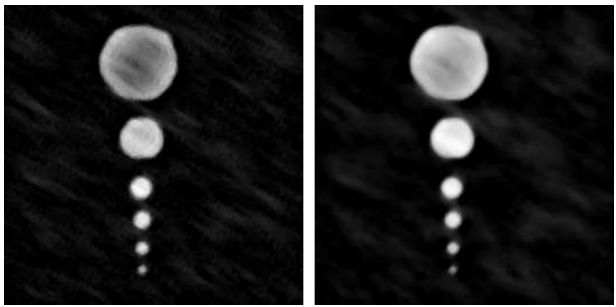}
\caption{\label{FigResult1b}
Left:  Reconstruction from the data in Fig.~\ref{FigTestimage1}, right, with chopping 
       amplitude $h=(5,3)$ using method (B) with $\tau=0.1$ and $100$ iterations combined 
       with median filtering (filter size $s=1$).
Right: Reconstruction after $30$ iterations of the cg-based method combined with median 
       filtering (filter size $s=3$ resp. $s=1$ in the last iteration).}
\end{center}
\end{figure}

Secondly, we consider reconstructions from the cg-based method and method (B). 
Results from method (A), which in general look more blurry,  are not presented here.

Fig. \ref{FigResult1b} left, shows the reconstruction from noisy data with 
chopping amplitude $h=(5,3)$ with method (B) combined with median filtering 
(filter size $s=1$). 
Fig. \ref{FigResult1b} right, shows the reconstruction using the cg-method with 
median filter ($s=3$ resp. $s=1$ in the last step). 

The computation times were 3.3 seconds for method (B) and 4.4 seconds for the cg-method.

In both reconstructions artificial 
structures arising from noise are removed by the median filtering 
(cf. Fig.~\ref{FigBertero1}).

\begin{figure}[h]\begin{center}
\includegraphics[width=0.45\textwidth]{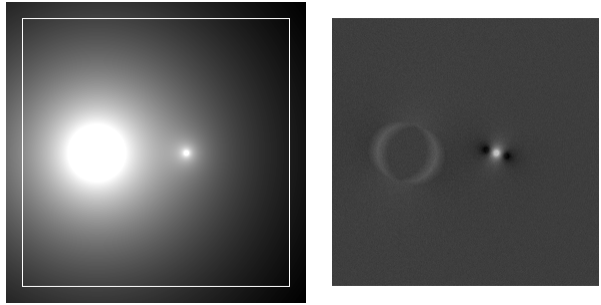}
\caption{\label{FigTestimage2}
Left: The second test image. 
Right: the corresponding noisy chopped and nodded image for $h=(10,3)$, 
       $\sigma_1= 0.002$ and $\sigma_2=  0.00001$.}
\end{center}
\end{figure}

\begin{figure}[h]\begin{center}
\includegraphics[width=0.45\textwidth]{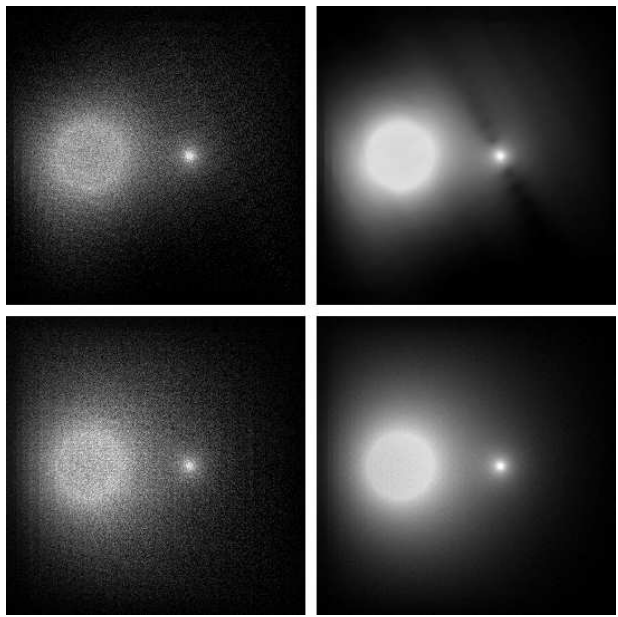}
\caption{\label{FigResult2}
Top left: Reconstruction with the cg-method after $40$ iterations (without median filtering). 
The chopping amplitude is $h=(7,14)$.
Top right:  $40$ iterations with the cg-method combined with median filtering with filter 
size $s=3$ ($s=1$ in the last iteration). 
Note that for comparison with the method of multiple chopped data sets with amplitudes 
$h= (10,3)$, $(0,7)$, $(10,10)$, $(7,14)$ and $(15,0)$ we depicted $h=(7,14)$ showing best 
results by independent reconstruction. 
Bottom left: Combined image using the multiple chopped data calculating the pointwise median after 
independent reconstruction.
Bottom right: Reconstruction using  multiple data sets and combining the results \emph{after each iteration} using the pointwise median.
}
\end{center}
\end{figure}

Further experiments have been performed with the test image shown in 
Fig.~\ref{FigTestimage2}, which contains an object with a wide halo extending
across the boundary of the chosen domain.

Fig.~\ref{FigTestimage2} (left) shows the signal $u$ on $\Omega_h$, the part in which the 
chopped data are provided is marked by the white rectangle.
We produced a multiple chopped data set with 5 different chopping amplitudes 
$(10,3)$, $(0,7)$, $(10,10)$, $(7,14)$ and  $(15,0)$.
The chopped data for $h=(10,3)$ on $\Omega$ is shown in Fig.~\ref{FigTestimage2} (right). 

We added noise of variance $\sigma_1=0.001$ and $\sigma_2=0.00001$. Note that for this test 
image the chopped data set contains very weak structures. The signal-to-noise-ratio is 
$40.0494$.
We compare different strategies for reconstruction from noisy data.
First of all we calculated the results of the cg-method with 40 iterations
for each single data set, applying the median filter after each iteration.
For a fair comparison we depict the data set showing the best results, which is that for 
$h=(7,14)$. Fig.~\ref{FigResult2} (top left) shows the result of the cg-based method 
without median filtering. Undesirable noise enhancement occurs.
Fig.~\ref{FigResult2} (top right) shows the reconstruction with the cg-method where in 
each iteration step a median filtering is performed 
($s=3$ resp. $s=1$ in the last step).

Fig.~\ref{FigResult2} (bottom left) shows the combined result of
independent reconstructions from the multiple data set, applying 
40 iterations of the cg-method without median filtering on each set and finally 
calculating the median in each data point. By this post-processing the artificial structures are weakened.

Fig.~\ref{FigResult2} (bottom right) shows the reconstruction with combining the 
different results after each cycle of cg-iterations by calculating the 
pointwise median.\\

The computation times were  about 0.9 seconds for the cg-method without median filtering and about 5.5 seconds with median filtering. 
The reconstruction with multiple chopped data sets took about 4.7 seconds
with independent reconstruction and finally averaging and 5 seconds with calculating the 
median after each iteration.

\begin{figure}[h]\begin{center}
\includegraphics[width=0.45\textwidth]{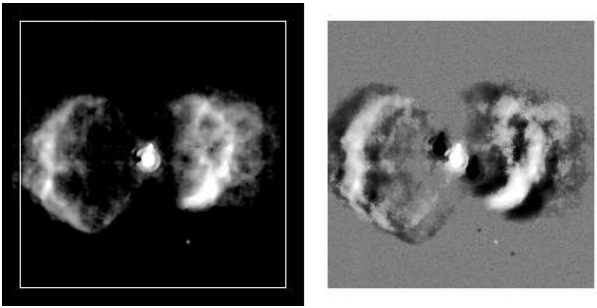}
\caption{\label{FigTestimage3}
Left: Third test image, planetary nebula Menzel III, Right: Artificial chopped data for $h=(14,10)$ including noise $\sigma_1= 5\;10^{-6}$ and $\sigma_2=10^{-6}$. 
}
\end{center}
\end{figure}

A third test image (Fig.~\ref{FigTestimage3}) is an observation of the planetary nebula 
Menzel III (the ant nebula).  The intensities stored in floating point precision show a 
high difference of intensities between the nebula and the central star. 
To visualize the data, the intensities were logarithmically scaled to gray values by 
applying the function $S(x)=\frac{log(x/(m\cdot max)+0.1)-log(0.1)}{log(1.1)-log(0.1)}$, 
where $x\ge 0$, $max$ is the maximal intensity in the image and $m$ is a scaling parameter, 
set to $m=0.01$ for the results presented here.
 
We simulated the chopping and nodding procedure including
artificial noise of variance $\sigma_1= 5\cdot 10^{-6}$ and $\sigma_2=10^{-6}$. 
We have chosen this noise variance corresponding to the weak intensities of the nebula in the range 
about $10^{-4}$.
We used small chopping amplitudes up to 14  pixels for testing. 
Note that when applying large chopping amplitudes the original image can be reconstructed 
by a few iteration steps. Then the amplification of noise is negligible.
The corresponding chopped data for $h=(14,10)$ is given in
Fig.~\ref{FigTestimage3} (right).
The signal-to-noise ratio is 42.9788.\\
\begin{figure}[h]\begin{center}
\includegraphics[width=0.45\textwidth]{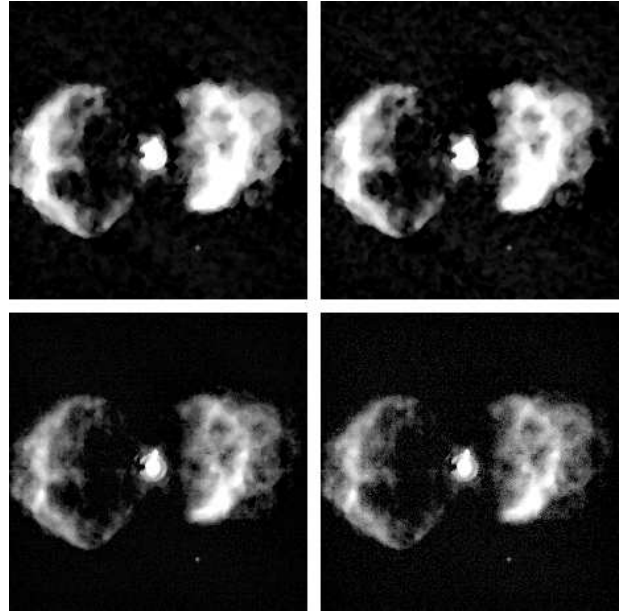}
\caption{\label{FigResult3}
Top left: Result of method (B) after $100$ iterations in combination with the median 
filtering with filter size $s=1$.
Top right: Result of the cg-based method after $20$ iterations in combination with 
median filtering with size $3$ resp. $1$ in the last iteration.
Bottom left: Result of a combined reconstruction with method (B) ($\tau=0.1$, 100 steps) 
using multiple data from chopping amplitudes $(14,10)$, $(14,0)$, $(14,14)$, $(10,0)$, and 
$(7,14)$. After each iteration step the reconstructed images are reinitialized with the 
pointwise median.
Bottom right: Result of a combined reconstruction (pointwise median) with the cg-method with 20 steps using the same multiple data set.
}
\end{center}
\end{figure}
Fig.~\ref{FigResult3} (top left) shows the result of method (B) after 100 iterations where 
at each iteration median filtering with $s=1$ has been used.
Fig.~\ref{FigResult3} (top right) shows the result of the cg-based method after 20 
iterations in combination with median filtering. In both cases some blurring effect caused by 
the median filter can be recognized.
Finally we used data obtained with chopping amplitudes 
$(14,10)$, $(14,0)$, $(14,14)$, $(10,0)$ and  $(10,14)$. Fig.~\ref{FigResult3} 
(bottom left) shows the reconstruction using method (B) on each data set and combining 
the results after each iteration step using the pointwise median as described in Sect.~\ref{multi}. 
In Fig.~\ref{FigResult3} (bottom right) we used the cg-method (20 iterations) where again 
the pointwise median is calculated after each iteration .

The computation times were about 2.5 seconds using method (B) with median filtering, 
0.9 seconds using the cg-method with median filtering and 2.8 resp. 1.6 seconds using  method (B) resp. 
the cg-based method on the five different chopped data sets and calculating the median after each iteration.

%------------------------------
\section{Summary and Conclusions\label{SectConclusion}}
%-------------------------------
In this paper we give an overview of Fourier-based and iterative schemes for 
reconstruction of images from chopped and nodded data. 
As an alternative to reconstruction methods documented in the literature we proposed an algorithm 
based on the conjugate-gradient(cg)-method which is faster converging. 

Experiments show that for the 'classical' reconstruction methods noise can severely 
affect the reconstruction process.
To prevent noise enhancement we propose to combine iterative methods 
with median filtering techniques (robust methods from statistics).
Filtering is performed in each iteration of the reconstruction process.

We produced satisfactory numerical results for three different test images with 
noisy chopping and nodding data. Noise artifacts are removed during reconstruction.

An alternative strategy to enhance the quality of reconstruction is to use multiple 
chopping amplitudes. The results of the reconstruction are combined 
by statistical methods after each iteration. This also provides a method robust against noise.

Since the method of multiple chopped data sets does not show any blurring effects, 
we propose to use this method if the required data are available. In the case of a single 
data set the methods with  median filter should be applied.

\begin{acknowledgements}
We thank Ulrich Kaeuffl from ESO in Garching for helpful discussions
about the topic of chopping and nodding and providing the Menzel III
 nebula data set for testing. For running the reconstruction algorithm the computer cluster of the 
HPC - Konsortium Innsbruck was used.
The work of F.L. is supported by the Tiroler Zukunftsstiftung.
The work of O.S. and S.S. is partly supported by the Austrian Science
Foundation,  Projects Y-123INF,  FSP Industrial Geometry 9201-N12 (subprojects 9207-N12 and 9203-N12), P15868  and the Uni Infrastructure II program. 
\end{acknowledgements}

%-----------------
%\FloatBarrier
\bibliography{astro.bib}
\bibliographystyle{aa}

\end{document}